\documentclass[%
preprint,
amsmath,amssymb,
aps
pra,
]{revtex4-2}

\usepackage{graphicx}% Include figure files
\usepackage{dcolumn}% Align table columns on decimal point
\usepackage{bm}% bold math
\usepackage{gensymb}
\usepackage{amsmath}
\usepackage{mathtools}
\usepackage{soul}
\usepackage{lineno,hyperref}
\usepackage{natbib}
\usepackage{amsfonts}
\usepackage{amssymb}
\usepackage{mathrsfs}
\usepackage{bbold}
\usepackage{latexsym}
\usepackage[table,xcdraw]{xcolor}
\usepackage[normalem]{ulem}
\usepackage{braket}
\usepackage{subfigure}
\usepackage{float}
\usepackage{comment}

\newcommand{\wag}[1]{\textcolor{blue}{#1}}

\begin{document}
	
	\preprint{APS/123-QED}
	
	\title{ Experimental investigation of discord  in spin-orbit X-states }% Force 
	
	%\title{Generation and characterization of two q-bits X-states using spin-orbit modes}% Force line breaks with \\
	
	%	\thanks{A footnote to the article title}%

	%\footnote{Corresponding author wagner.balthazar@ifrj.edu.br.com}}
	%	\altaffiliation[Also at ]{Physics Department, XYZ University.}%Lines break automatically or can be forced with \\
	\author{ V. S. Lamego$^{1,2}$, D.G. Braga$^{1,2}$, W. F. Balthazar$^{2,3}$, and J. A. O. Huguenin$^{1,2}$}%
	\email{jose$_$huguenin@id.uff.br}
	
	\affiliation{%
		$^1$ Programa de Pós-graduação em Física - Instituto de F\'{\i}sica - Universidade Federal Fluminense - Niter\'{o}i - RJ - Brasil\\
		$^2$	Instituto de Ciencias Exatas - Universidade Federal Fluminense - Volta Redonda - RJ - Brasil \\
		$^3$	Instituto Federal de Educação, Ciência e Tecnologia do Rio de Janeiro - IFRJ\\
	}%
	
	%	\collaboration{MUSO Collaboration}%\noaffiliation
	
	%	\author{Charlie Author}
	%	\homepage{http://www.Second.institution.edu/~Charlie.Author}
	%	\affiliation{
	%		Second institution and/or address\\
	%		This line break forced% with \\
	%	}%
	%	\affiliation{
	%		Third institution, the second for Charlie Author
	%	}%
	%	\author{Delta Author}
	%	\affiliation{%
	%		Authors' institution and/or address\\
	%		This line break forced with \textbackslash\textbackslash
	%	}%
	
	%	\collaboration{CLEO Collaboration}%\noaffiliation
	
	\date{\today}% It is always \today, today,
	
\begin{abstract}		
We perform an experimental investigation of Quantum Discord with Spin-Orbit X-states. These states are prepared through the incoherent superposition of different laser beans, where a two-level system is encoded in polarization and the first-order Hermitian-Gaussian modes, as proposed in Phys. Rev. A 103,0022411 (2022). We characterize different classes of spin-orbit X-states by performing an all-optical tomography for polarization and first-order transverse mode degrees of freedom.
We also perform a study on the dependence of Discord with respect to the Fidelity of spin-orbit modes, revealing that Discord is very sensitive to incoherent noise. Our experimental results align with the theoretical predictions of Quantum Discord when accounting for the effect of fidelity. These results reinforce the spin-orbit modes as an important platform for quantum information processing. On the other hand, the need of astigmatic optical elements to implement unitary operations required by quantum information protocols implies in a loss of fidelity and quantum correlations. Alternative methods to manipulate spin-orbit modes are welcome.
\end{abstract}
	
	%Use showkeys class option if keyword
 
	\maketitle
	
	%\tableofcontents
	
	\section{Introduction}

Quantum Discord (QD) \cite{QD_zurek} is an important resource for quantum computation and quantum information \cite{QD_brodu}, exploring a diverse range of scenarios where the relation of classical correlations and quantum mutual information is pivotal. These scenarios include applications such as remote state preparation for quantum communication \cite{dakic}, critical systems \cite{sarandy} to Mixed-State Metrology \cite{modi}, spin models \cite{Huang}, and multiqubit systems \cite{li}. In the discord dynamics scenario, investigations have been conducted on two-qubit systems in non-Markovian scenarios \cite{fanchini, alipour}, as well as in the context of photonic crystals \cite{wang} and two-photon states \cite{lemr}.

Considering some specific cases of mixed states, for a given amount of classical correlation, it is important to maximize quantum discord. This task was achieved by using unbalanced Bell state, giving rise to the Maximally Discordant Mixed State (MDMS) \cite{Galve}. Such states were generated by using dissipative schemes \cite{xxli}. Two separable qubit presented discord quantified by local operations \cite{gharibian}. Lately, it was demonstrated necessary conditions for obtainment of a maximally discordant separable two qubit X-states \cite{rana}. X-states scenario is, indeed, a fertile environment for discord studies \cite{eberly, ara, chen}.

%%%%  spin-orbit

Among the main platforms for implementing protocols and studies on quantum information and quantum computation, the use of structured light stands out. By associating polarization degree of freedom (DoF) and first-order transverse mode, we can prepare the well-known Spin-Orbit (SO) mode \cite{topo}. The quantum-classical analogy, presented by a laser beam prepared in nonseparable spin-orbit modes, has been investigated in experiments that showed violation of Bell inequality pointing out the similar structure of nonseparable SO modes with entangled states  \cite{carol, kagawala, qian}. By adding path DoF, a tripartite-like was prepared and it was observed violation of Mermim's inequalities \cite{tripartite}. These SO modes have also been used to study other correlations, such as contextuality \cite {contex}. Genuine entanglement between different DoF of a single photon emerges from quantization of SO modes \cite{pereira}.

In a similar way, this platform has been employed to emulate various quantum information and quantum computation protocols, including quantum key distribution without referential frame \cite{qkdspinorbit}, teleport protocol between DoF of light \cite{telezela}, and quantum channels \cite{qchannel}. Fundamental aspects have been explored as environment-induced entanglement \cite{environE} and non-Markovian signatures \cite{nonMarkov}. A seminal contribution was made through the optical simulation of quantum thermal machine \cite{quantumthermal}. We cannot forget the various implementations of logic gates exploiting structured light \cite{qg1, qg2, qg3}.

%%% discord spin-orbit modes

The study of QD presented in this paper is grounded in the preparation of X-states using SO modes \cite{Xstate}. Different optical circuits were proposed for the preparation of different classes of states, including Werner states and other incoherent superpositions. A linear optical circuit for tomography of spin-orbit DoF was introduced. The preparation and characterization were computationally performed, and the calculated QD exhibited a remarkable agreement with theoretical expectations. Maximally Discordant Mixed States were proposed \cite{SpinOrbitMDS}, necessitating the preparation of an unbalanced non-separable mode, recently achieved experimentally \cite{PNSM}. However, to the best of our knowledge, no experimental studies of QD of spin-orbit X-states have been performed. 

%%%%   IN THIS WORK

In this work, we present the measurement of discord of different classes of spin-orbit X-states. We perform spin-orbit tomography. The effect of a non-faithful reconstruction of those of the states in Discord is investigated. The paper is organized as follows. In section II we present a theoretical review, present discord calculation, and investigate spin-orbit modes. In Section III we present a study of the dependence of QD with fidelity. Optical circuits for X-state preparation and the optical circuit to perform spin-orbit tomography, are presented in Section IV. Results are presented and discussed in Section V. Finally,  concluding remarks are presented in Section VI.

 %%%%%% OLD TEXT  TO BE REMOVED %%%%%%%%%

\section{Theoretical review} \label{sectionDisc}

\subsection{Discord}

Quantum Discord (QD)\cite{QD_zurek} is widely discussed in the literature, finding applications in the development of quantum technologies and the study of the fundamentals of quantum mechanics. In technological contexts, QD serves as a quantum resource, enabling the exploration of quantum processing capabilities necessary to achieve advantages in specific tasks \cite{remdisc,cripdisc, compdisc}. Additionally, QD serves as an indicator of the system's 'quantumness', distinguishing quantum correlations from classical ones \cite{sarandy, corredisc}. It is important to note the abundance of definitions for QD due to the complexity of the quantum realm, such as Entropic Discord and Geometric Quantum Discord \cite{reviewdisc}.  For instance, for a bipartite system described as a density operator $\rho_{AB}$, the entropic quantum discord is defined as
\begin{equation}
\mathcal{Q}(\rho_{AB}):=\mathcal{I}(\rho_{AB})  -\mathcal{C}_C(\rho_{AB}),
\label{discordia}
\end{equation}
where $\mathcal{I}(\rho_{AB})$ is the usual quantum mutual information and $\mathcal{C}_C(\rho_{AB})$ is the classical correlation given by
\begin{equation}
	\mathcal{C}_C(\rho_{AB})=\mathcal{S}(\rho_A)-\min_{ \{B_i\}}[\mathcal{S}(\rho_{AB}|\{B_i\})],
	\label{classicalcorrelation}
	\end{equation}
being $\mathcal{S}(\rho_A)$ the von neumann entropy of subsystem A and $\mathcal{S}(\rho|\{B_i\})$ a measurement-induced conditional entropy obtained from
\begin{equation}
	\mathcal{S}(\rho_{AB}|\{B_i\})=p_0S(\rho_0)+p_1S(\rho_1).
	\label{qconditional entropy}
	\end{equation}

Note that $p_0$ and $p_1$ are the probabilities of finding the system at the states $\rho_0$ and $\rho_1$  after performing a local Von Neumann measurement  $\{B_i\}$  on the subsystem B. It is evident from Eq. (\ref{classicalcorrelation}) the necessity of a minimization procedure over all sets of Von Neumann measurement operators $\{B_i\}$ that can be addressed via numerical methods. It's relevant to emphasize that equation \ref{discordia} captures the discrepancy of two possible ways to quantize classical mutual information, as explained in reference \cite{luo2008}. In this paper, we compute entropic QD by a computational approach:  classical correlation is maximized using scanning over all possible sets of projective measurement operators $\{B_i\}$ and then find the one that maximizes quantity $\mathcal{C}_C(\rho_{AB})$. Therefore, precise knowledge of the system's density operator is critical for discord computation.

 \subsection{Studied mixed spin-orbit X-states }

Spin-orbit (SO) modes can be understood as an integrated description of the electromagnetic field by polarization and transverse modes DoF \cite{topo}. The most general SO modes are linear polarization $\hat{e}$ and first-order HG modes, which can be written as

\begin{equation}
\begin{split}
    \vec{E}_{SO}(\vec{r}) = & \, \alpha HG_{10}(x,y)\hat{e}_H + \beta HG_{10}(x,y)\hat{e}_V \\
    & + \gamma HG_{01}(x,y)\hat{e}_H + \delta HG_{01}(x,y)\hat{e}_V,
\end{split}
\label{spinorbitGEN}
\end{equation}

\noindent where the Greek coefficients are complex numbers that obey the relation $|\alpha|^2 + |\beta|^2+ |\gamma|^2+ |\delta|^2=1$. When such a basis is used for the quantization of the electromagnetic field, we can write the quantum state of the electromagnetic field as 
\begin{equation}
\begin{aligned}
    \ket{\psi_{SO}}=& a_{Hh} \ket{Hh}  +  a_{Hv} \ket{Hv} + a_{Vh} \ket{Vh} \\& +  a_{Vv} \ket{Vv},
\end{aligned}
\label{spinorbitQSTATE}
\end{equation}
where we used the notation $\hat{e_H}\equiv H$, $\hat{e_V}\equiv V$ for polarization DoF, and $HG_{01}(\vec{r})\equiv h$, $HG_{10}(\vec{r})\equiv v$ for transverse mode DoF. The probabilities amplitudes $a_{ij}$ ($i-H,V,~j=h,v$) are normalized.

Following the proposal of Ref. \cite{Xstate}, the emulation of quantum mixed states can be accomplished by combining independent laser beams, with each beam encoding a pure state of two qubits. Once the lasers are independent, there is no well-defined phase relation between them and that is the reason why they do not interfere. So the measurement outcomes will be the sum of the contribution of intensity of each beam, providing the statistics of a mixed state. 

The mixed modes experimentally investigate are the following:

\begin{equation}
    \rho_1 = c\ket{\phi^+}\bra{\phi^+} + \left(1-c\right)\ket{Vv}\bra{Vv},
    \label{eq6}
\end{equation}
\begin{equation}
    \rho_2 = c\ket{\phi^+}\bra{\phi^+} + \left(1-c\right)\ket{\psi^-}\bra{\psi^-},
    \label{eq7}
\end{equation}
 \begin{equation}
    \rho_3 = \frac{1}{3}\left(c\ket{Vv}\bra{Vv} + \left(1-c\right)\ket{Hh}\bra{Hh} + \ket{\psi^-}\bra{\psi^-}\right),
    \label{eq8}
\end{equation}

\noindent where c $\in [0,1]$. Here, $\ket{\phi^+} = \frac{1}{\sqrt{2}}\left(\ket{Hh}+ \ket{Vv}\right)$ and $\ket{\psi^-} = \frac{1}{\sqrt{2}}\left(\ket{Hv}- \ket{Vh}\right)$ are the maximally nonseparable mode, analogue of Bell states \cite{carol}.

Once we need the density matrix to compute Quantum Discord, we must implement a two-qubit quantum-like state tomography process to estimate the density matrix. In our optical experimental setup, involving preparation and tomography, there are many typical sources of error, including imperfections in optical elements, misalignment, and, mainly, imperfect mode matching for the mode converter, among others. For spin-orbit modes, the measurement in $y$-basis needs astigmatic optical elements, specifically the mode converter, which serves as a significant source of error in density matrix reconstruction. In this sense, experimental errors can impose limitations on fidelity in both the state preparation and tomography steps.

For this reason, in the next section, we will present a study on the role of Fidelity of our studied modes in the discord calculation.

  \section{Discord and fidelity}

Taking into account the experimental errors during the realization of the experiment, it is crucial to consider them in the Quantum Discord (QD) calculation. Following a practical approach, we present a study examining the relationship between the effects of lower fidelity and the resulting discord (QD).

One potential model involves decoherence effects that results in the intended state becoming a combination of the considered density operator and an identity term. Then, to study this effect, we insert an identity term with a controllable weight in states $\rho_1,~\rho_2,$ and $\rho_3$ to observe the effect of fidelity's degradation on the Discord's curve. A perturbed state $\rho^\prime$ is defined as
\begin{equation}
    \rho^\prime = (1-\alpha)\rho + \alpha I,  \label{rhopertub}
\end{equation}
where $\alpha$ is the weight of identity term. We used $\alpha=0;~0.1;~0.2;~0.3$.

\begin{figure}[ht!]
	\centering
		\includegraphics[scale=0.4,trim=0cm 0cm 0cm 0cm, clip=true]{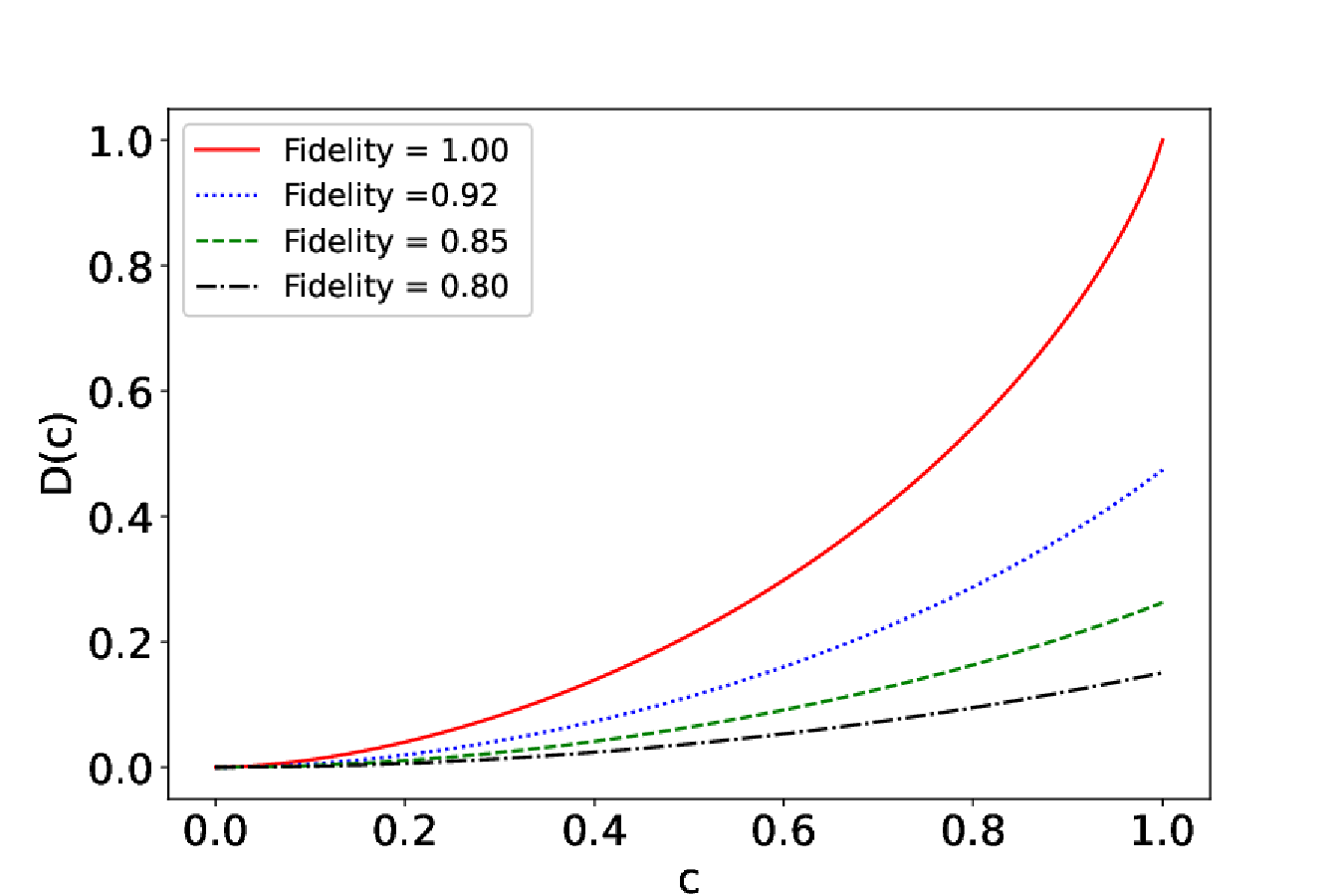}
		\caption{ Discord in function of the weight c for state $\rho_1$ combined to identity. Mean fidelity $=1.00;~0.92;~0.85;~0.80$ corresponds to $\alpha=0;~0.1;~0.2;~0.3$, respectively.} 
		\label{discordStudy1}
	\end{figure}

Figure \ref{discordStudy1} presents the results for Discord of $\rho_1$. The theoretical expectation is observed for Fidelity $= 1.00$  (solid line - red online). 
The Mean fidelity $F$ decreases as $\alpha$ increases, which leads to a degradation of discord as we can see for the studied cases of $F=0.92$ ($\alpha=0.1$ - dotted line - blue online),  $F=0.85$ ($\alpha=0.2$ - dached line - green online), and ses of $F=0.80$ ($\alpha=0.4$ - dached-dotted line - black online). For $c>0.1$ we can see that Discord degrades more and more sharply as Fidelity decreases. The pure Bell State ($c=1$) shows more sensitivity to the loss of Fidelity. 

\begin{figure}[ht!]
	\centering
		\includegraphics[scale=0.4,trim=0cm 0cm 0cm 0cm, clip=true]{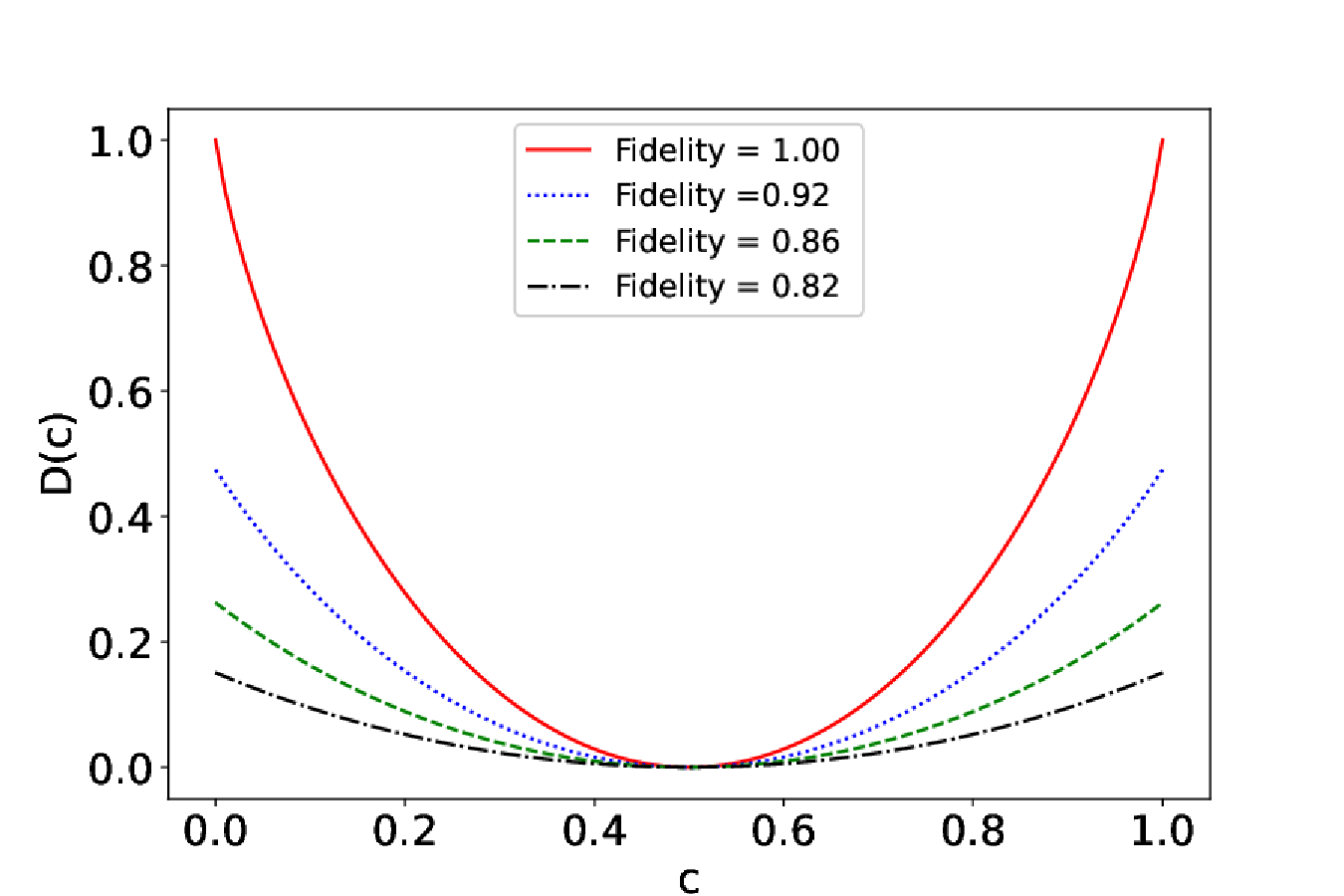}
		\caption{ Discord in function of the weight c for state $\rho_2$ combined to identity. Mean fidelity $=1.00;~0.92;~0.86;~0.82$ corresponds to $\alpha=0;~0.1;~0.2;~0.3$, respectively.} 
		\label{discordStudy2}
	\end{figure}

Figure \ref{discordStudy2} presents the results for $\rho_2$ (mixing of two Bell-like modes).  For $c=0.5$, Discord is null as expected, and the results show that decrease in Fidelity for $0.4 < c < 0.6$. As $c$ approaches $0$ or $1$, Discord degrades when Fidelity decreases. Again, pure Bell-like modes are the more affected by noises. 

\begin{figure}[ht!]
	\centering
		\includegraphics[scale=0.4,trim=0cm 0cm 0cm 0cm, clip=true]{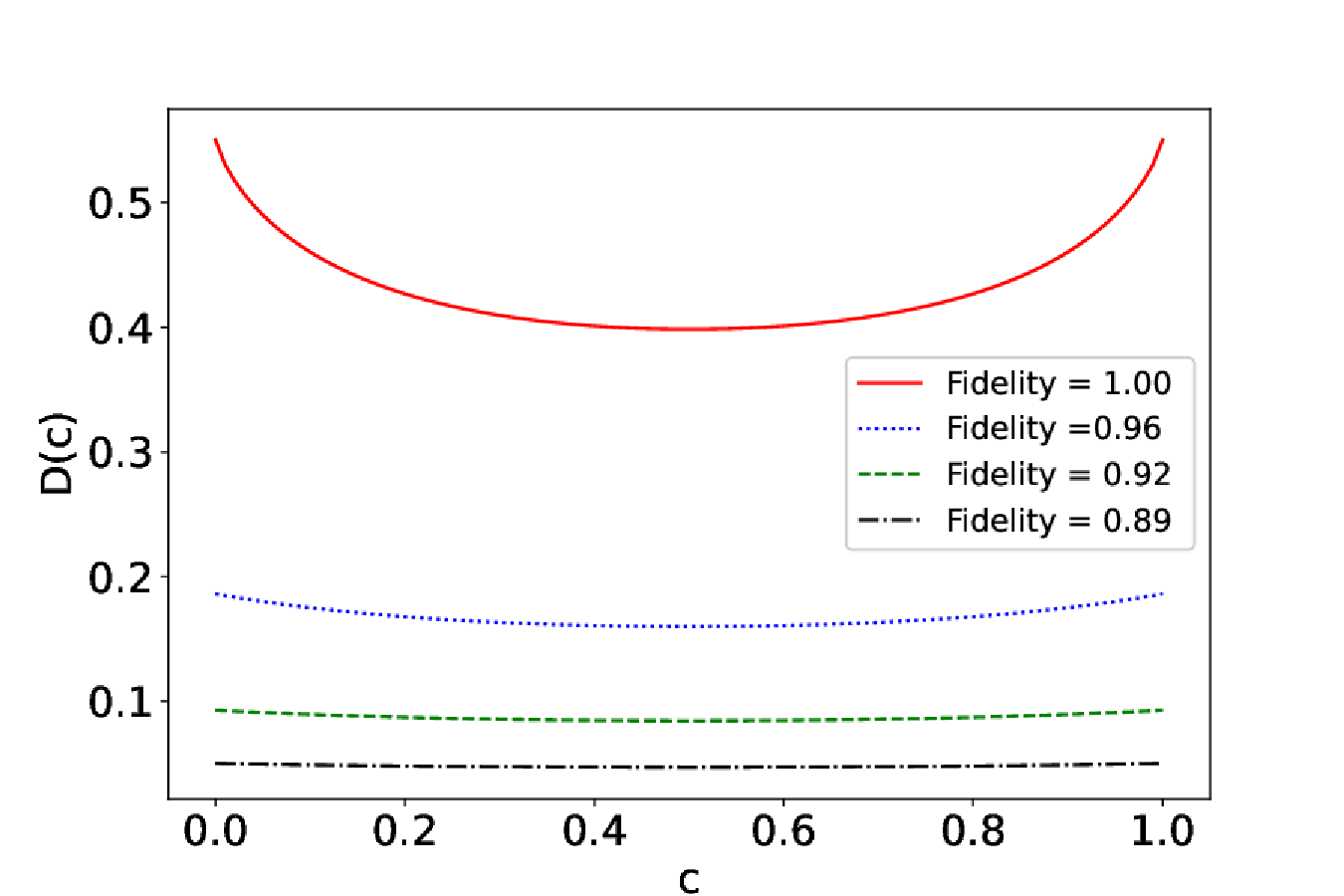}
		\caption{ Discord in function of the weight c for state $\rho_3$ combined to identity. Mean fidelity $=1.00;~0.96;~0.92;~0.89$ corresponds to $\alpha=0;~0.1;~0.2;~0.3$, respectively.} 
		\label{discordStudy3}
	\end{figure}

Finally, for $\rho_3$, where we have a fixed term $ \frac{1}{3}\ket{\psi^+}\bra{\psi^+}$, that guarantee Discord never vanishes, Figure \ref{discordStudy3} shows we have a more degrading behavior of Discord concerning Fidelity. Note that for $F=0.9$ Discord partially vanishes independently of $c$'s values.

Despite the high fidelity is observed an appreciable deviation as the fidelity decreases, being the effect more pronounced for coherent states. This analysis will help us to interpret our experimental results which will be presented later. As will be shown in section V, this model captures with good approximation the experimental data behavior. Provided these considerations, we can advance to the following section to present the optical setup necessary to prepare and characterize the states considered in this paper.

\section{Experiment}
	
\subsection{ State preparation } 

\begin{figure}[H]
		\centering
		\includegraphics[scale=.35,trim=0cm 0cm 0cm 0cm, clip=true]{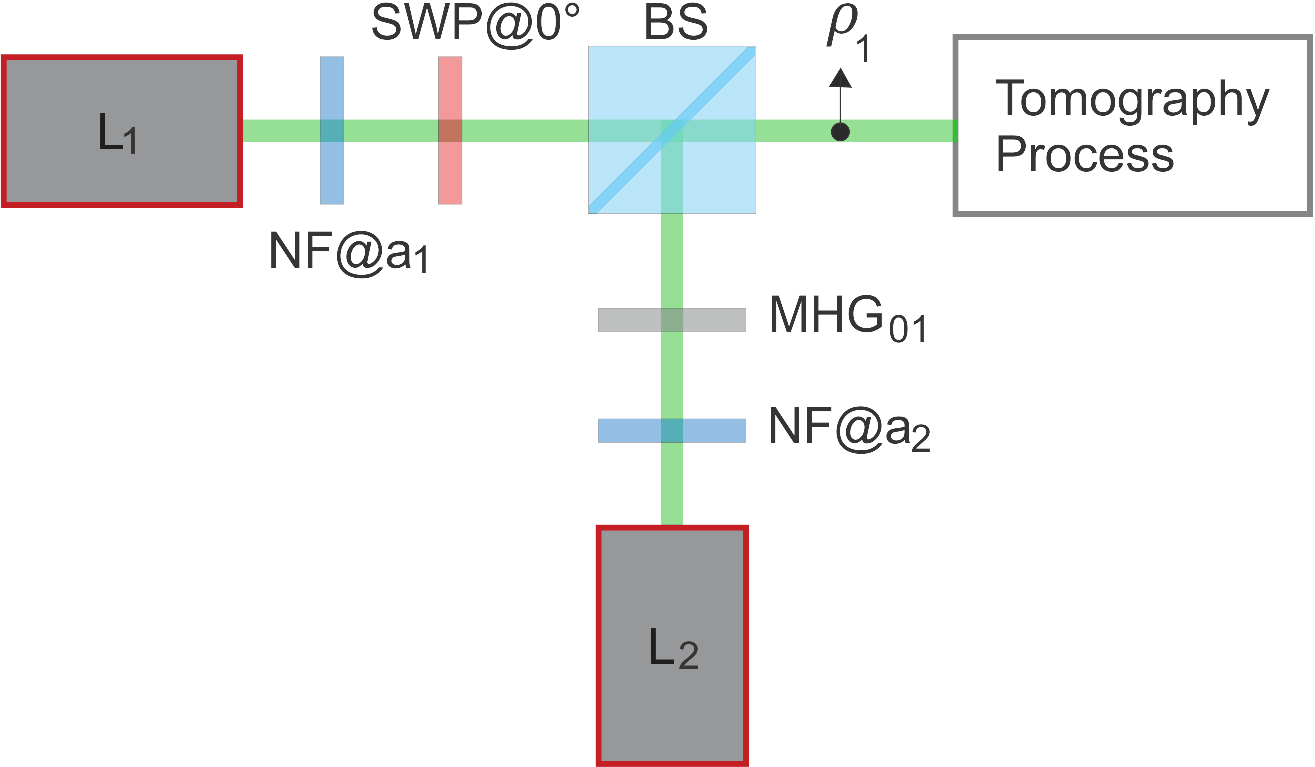}
		\caption{Experimental setup to generate $\rho_1$ state. L$_i$ ($i=1,2$) stands for lasers. M$_{HG}$ stands for holographic mask. NF stands for variable neutral filter.  BS stands for beam splitter.SWP stands for s-waveplate.} 
		\label{OpticalCircuit1a}
\end{figure}

\begin{figure}[H]
		\centering
		\includegraphics[scale=.35,trim=0cm 0cm 0cm 0cm, clip=true]{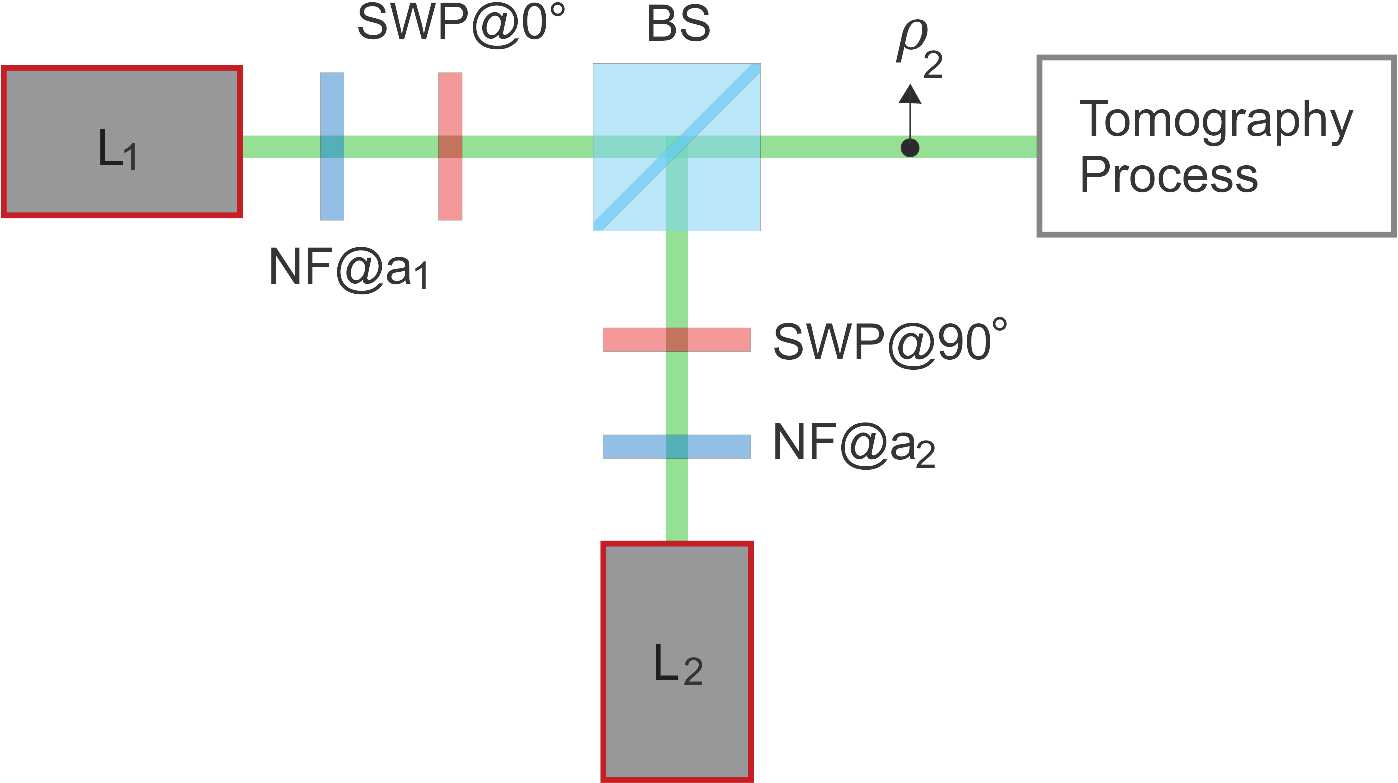}
		\caption{Experimental setup to generate $\rho_2$ state. L$_i$ ($i=1,2$) stands for lasers. PBS stands for polarized beam splitter. BS stands for beam splitter. HWP stands for half waveplate. SWP stands for s-waveplate.} 
		\label{OpticalCircuit1b}
\end{figure}

\begin{figure}[H]
		\centering
		\includegraphics[scale=.3,trim=0cm 0cm 0cm 0cm, clip=true]{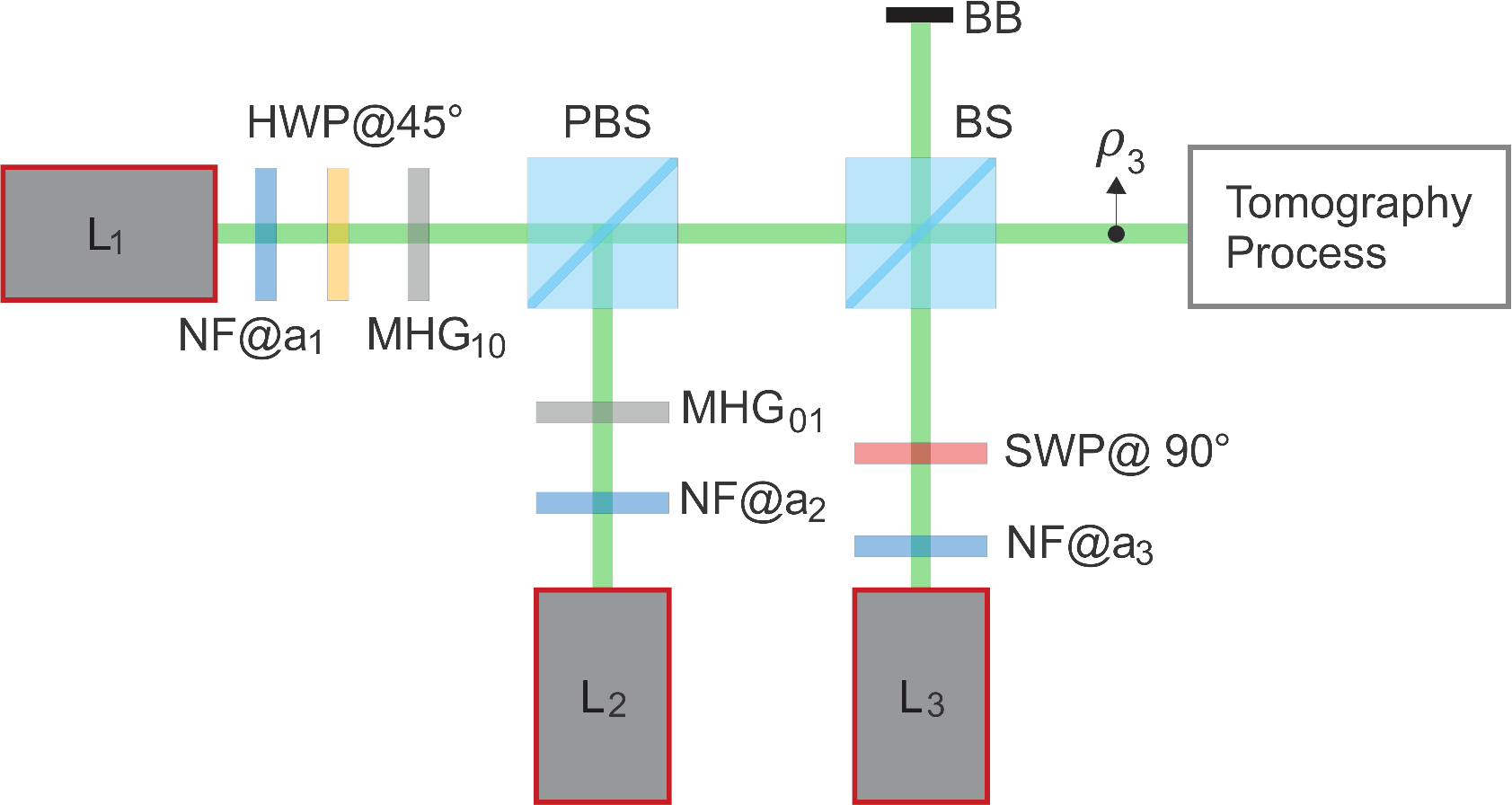}
		\caption{Experimental setup to generate $\rho_3$ state. L$_i$ ($i=1,2,3$) stands for lasers.M$_{HG}$ stands for holographic mask. NF stands for variable neutral filter.  BS stands for beam splitter. SWP stands for s-waveplate. BB stands for beam block.} 
		\label{OpticalCircuit1c}
\end{figure}

 To prepare the states $\rho_1$, $\rho_2$ and $\rho_3$, given, respectively, by Equations (\ref{eq6}), (\ref{eq7}) and (\ref{eq8}), we use the circuits represented in the figures \ref{OpticalCircuit1a},  \ref{OpticalCircuit1b},  and \ref{OpticalCircuit1c}, respectively. All those optical circuits work in the same way: we prepare ensembles of pure states codified on independent beams that are combined resulting on a mixed state.
  
The optical circuit for preparing the state $\rho_1$, given by Eq. (\ref{eq6}), is shown in Figure \ref{OpticalCircuit1a}. An S-wave plate with the axis on angle $@0$°  ($SWP@0^\circ$) with the horizontal acts converting a vertically polarized input laser beam (L$_1$) into a maximally nonseparable spin-orbit mode $\ket{\phi^+}$. The  neutral filter $NF@a_1$ with adjustable transmittance ($T_{a_1}$) is used to control the intensity of mode $\ket{\phi^+}$. 

On the other branch, an independently operated laser beam (L$_2$), vertically polarized, illuminates a holographic mask $M_{HG_{01}}$ to prepare the $HG_{01} \equiv v$ mode. The neutral filter $NF@a_2$ with adjustable transmittance ($T_{a_2}$) is used to control the intensity of mode $\ket{Vv}$.

Both independent pure modes are superposed in a $50/50$ beam splitter ($BS$) to produce mixed spin-orbit modes  $\rho_1$. By controlling the transmittance $T_{a_1}$ and $T_{a_2}$ of the neutral filters, the suit intensity proportions are adjusted to get the desired mixed state. To reproduce the mode class given by Eq.\ref{eq6} we set $T_{a_1}=c$ and $T_{a_2}= 1- c$.

For $\rho_2$ mode, given by Eq.\ref{eq7},  the preparation requires the combination of two independent maximally nonseparable modes, $\ket{\phi^+}$ and $\ket{\psi^-}$. Then, as shown in Figure \ref{OpticalCircuit1b}, a vertically polarized laser beam is directed through a $SWP@0^\circ$ to prepare the mode $\ket{\phi^+}$. The neutral filter $NF@a_1$ controls the intensity using its transmittance $T_{a_1}$.  A second vertically polarized laser (L2) has intensity controlled by the neutral filter $NF@a_2$ with transmittance $T_{a_2}$, which is directed through  a second S-wave plate at $90^\circ$ ($SWP@90^\circ$) to prepare the mode $\ket{\psi^-}$. A $50/50$ BS is used do superpose the mixed mode $\rho_2$ also by setting $T_{a_1}=c$ and $T_{a_2}=1-c$.

Finally, to prepare the mode $\rho_3$ given by Eq.\ref{eq8}, we used the optical circuit presented in Figure \ref{OpticalCircuit1c}. Note that the modes class $\rho_3$ is composed of a fixed fraction ($1/3$) of the maximally nonseparable mode $\ket{\psi^-}$, and the remains $2/3$ of the intensity is composed by a variable convex combination of separable modes $\ket{Hh}$ and $\ket{Vv}$.

Then, A laser (L$_1$) has its vertical polarization rotated $90^\circ$ by the HWP$@45^\circ$ and the Mask$HG_{10}$ produces the $h$ HG mode which completes the $\ket{Hh}$ mode preparation. Its intensity is controlled by the neutral filter $NF@a_1$ with transmittance $T_{a_1}$. A second laser (L2) prepares the mode $\ket{Vv}$ by using the Mask$HG_{01}$ and has its intensity controlled by the neutral filter $NF@a_2$ with transmittance $T_{a_2}$. A third laser (L3) vertically polarized shines the $SWP@90^\circ$ produce the maximally nonseparable mode $\ket{\psi^-}$, which has its intensity controlled by $NF@a_3$ with transmittance $T_{a_3}$. The there modes are superposed in $50/50$ BS and the transmittance are adjusted as $T_{a_3}=1/3$ of the total intensity, $T_{a_1}=2c/3$ and $T_{a_2}=2(1-c)/3$.

 The prepared state can be characterized by a tomography process, enabling us to reconstruct the state just by accessing statistics obtained by a set of projective measurements on different basis, as mentioned earlier.

\subsection{ Optical  circuits for spin-orbit Tomography}

\begin{figure}[H]
    \centering
    \includegraphics[scale=0.275,trim=0cm 0cm 0cm 0cm, clip=true]{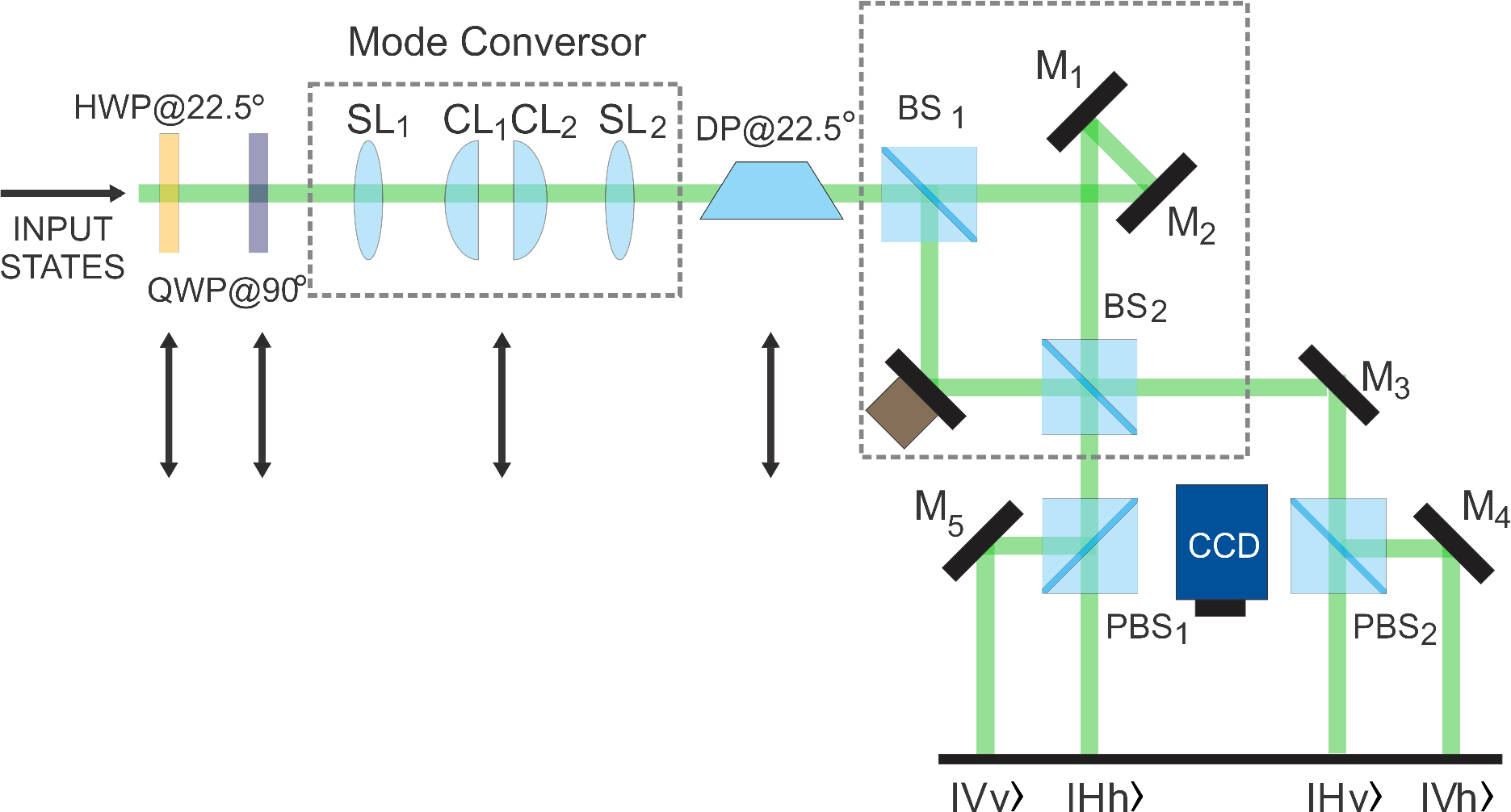}
    \caption{ Experimental setup to perform tomography. CL stands for cylindrical lenses. DP stands for Dove Prism. HWP stands for half waveplate. QWP stands for quarter waveplate. MZIM stands for Mach-Zehnder interferometer with one additional mirror. M stands for mirror. M,PZT stands for mirror with a piezoelectric ceramic. PBS stands for polarized beam splitter. BS stands for beam splitter.}
    \label{tomo}
\end{figure}

The spin-orbit mode tomography process can be implemented optically by the setup represented in Figure \ref{tomo}, following \cite{Xstate}. 
This setup enables state reconstruction through a series of intensity measurements, which are used to calculate all Stokes parameters, $S_{ij}$ \cite{Altepeter}, where \wag{~} $i,~j=0, 1, 2, 3$. The labels $i$ and $j$ are associated with polarization and transverse mode, respectively.

 The MZIM (Mach-Zehnder Interferometer with an additional Mirror) associated with PBS$_1$ and PBS$_2$ measure the two qubit computational basis $\{Hh,~Hv,~Vh,~Vv\}$. The parity selection performed by the MZIM associated with PBS's gives us four intensity outputs $I_{ij}$ related to the projections in each computational basis component. The outputs shine a screen that is imaged by a monochromatic CCD camera and we have an image with the record of the four intensities simultaneously which able us to calculate the probability of each basis component as
 \begin{equation}
     P_{i,j} = \frac{I_{i,j}}{I_T},
 \end{equation}
where $I_{i,j}$ ($i=H,V;~j=h,v$) is the gray-scale intensity of component $ij$ given by the CCD and $I_T$ is the total intensity obtained by the sum of the four outputs.

The measurement in the rotated basis of polarization DoF is performed by adding alternately in the input state path a half-wave plate at $22.5^\circ$ ($HWP@22.5^\circ$) and a quarter wave plate at $90^\circ$ ($QWP@90^\circ$). To measure the diagonal/antidiagonal ($D/A$) basis we include only the $HWP@22.5^\circ$. To measure the right/left circular basis ($R/L$) we used both $HWP@22.5^\circ$ and $QWP.@90^\circ$. Of course, always associated with MZIM and PBS units.

For the $d/a$ and $r/l$ basis for transverse mode DoF, a Dove Prism with its base rotated by $22.5^\circ$ plays an analog role of $HWP@22.5^\circ$ for $h$ and $v$  $HG$ modes. While a $\pi/2$ mode converter composed of two cylindrical lenses plays the role of $QWP.@90^\circ$. It is worth mentioning that, to satisfy the mode matching requirement for $\pi/2$ mode converter it is necessary to associate a pair of spherical lenses, denoted as $SL_1$ and $SL_2$ as shown in Figure \ref{tomo}. This is a very sensitive point and a source of the most pronounced experimental error. 

A detailed description of all tomographic measurement sets for each Stokes parameter for spin-orbit tomography can be found in the proposal of Ref. \cite{Xstate}.

It is worth mentioning that we control the waist of all lasers with a set of lenses for each laser, since we need to superimpose the modes of each independent laser to prepare the states. Most critically, for tomography, all modes will have to comply with the mode matching conditions in the cylindrical lens telescope and must be aligned simultaneously in the MZIM. These lens assemblies are not shown in the experimental setups. Such as control was fundamental for the measurement.

\section{Experimental Results}

In this section, we present the experimental results for spin-orbit mixed states tomography, as expressed by Eqs. (\ref{eq6}), (\ref{eq7}), and (\ref{eq8}). The experimental findings for quantum discord, considering fidelity, are presented.

\subsection{Recovered spin-orbit modes density matrix}  

Let us first present the results for state $\rho_1$. We will present the obtained images for $c=0;~c=0.5;$ and $c=1$ 
 
\begin{figure}[H]
		\centering		\includegraphics[scale=0.48,trim=0cm 0cm 0cm 0cm, clip=true]{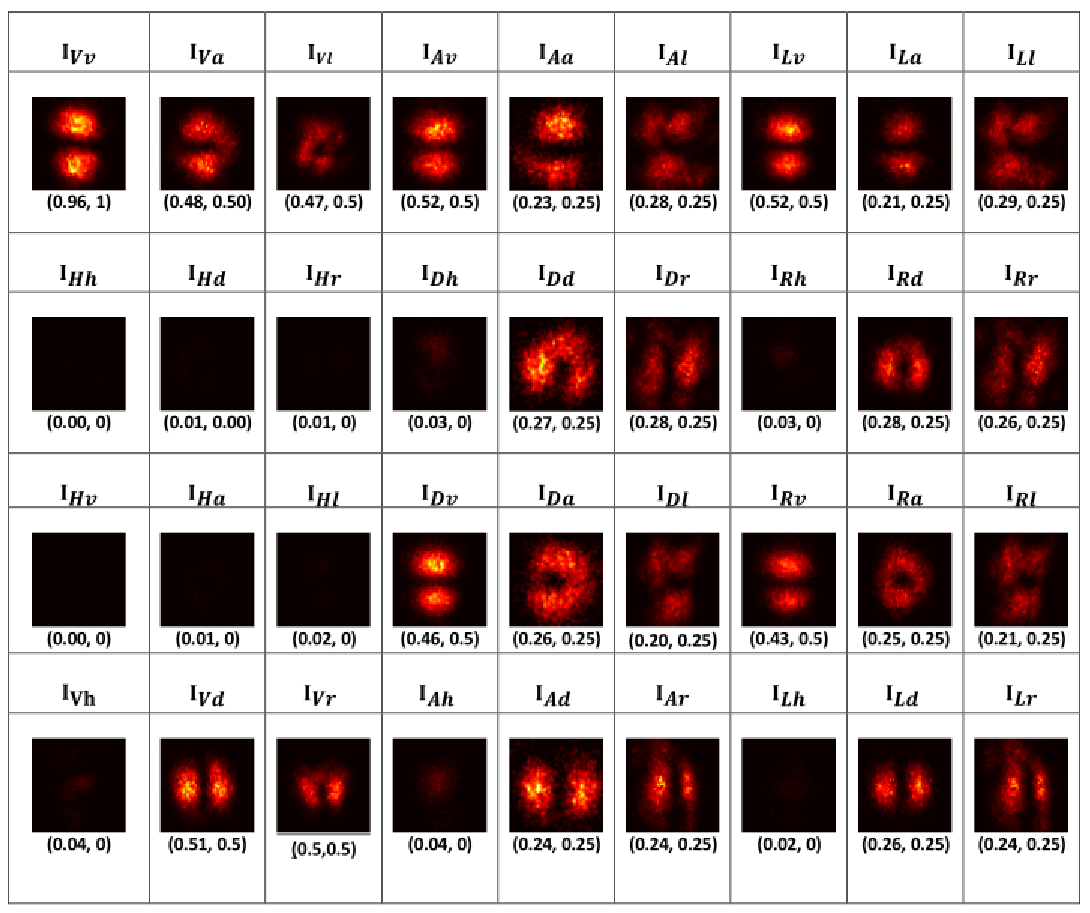}
		\caption{  Recorded images (false color) for $\rho_1$ in the case $c=0$, corresponding to the separable mode $\ket{Vv}$ for tomographic measurements. Within the parenthesis, we provide the experimental ($I{exp}$) and the theoretical ($I{theo}$) normalized intensities. The pair $(I_{exp}, I_{Theo})$ corresponds to experimental and theoretical normalized values, respectively, for the correspondent basis element.}
		\label{table1}
	\end{figure}

Figure \ref{table1} presents the recorded images (false color) for the tomographic measurement for $c=0$ that corresponds to the separable mode $\ket{Vv}$ for each set of basis. The image labeled as $I_{Vv}$ is the intensity recorded in the output for the component $\ket{Vv}$, corresponding to the set of the computational basis for Polarization $\{H,V\}$ and Transverse mode $\{h,v\}$. The image labeled as $I_{La}$ corresponds to measurement on a rotated basis, that is the intensity recorded in the output for the component $\ket{La}$, corresponding to the mode with left circular polarization ($\ket{L}$), corresponding to component $L$ measured with a set of rotated basis Right/Left ($\{R,L\}$) for Polarization and the Transverse Mode $~\ket{-}~$ corresponding to the component $"a"$ of the measurement in the diagonal/anti-diagonal basis $\{d,a\}$. Then, all labels indicate one result of the 36 combined measurements in the three bases for each DoF required for tomography \cite{Xstate}. 

The parenthesis below each image presents experimental and theoretical normalized intensities, in pairs $(I_{exp}, I_{Theo})$, respectively, for the correspondent basis element.

  Although the values are generally close, certain deviations can be attributed to previously mentioned experimental errors. As the number of optical elements increases, these deviations tend to amplify, as expected. Particularly, measurements on the basis set $\{Rr, Rl, Lr, Ll\}$ exhibit more deviations as it requires more optical elements. Additionally, these measurements are notably affected by astigmatic elements of the mode converter and the Dove prism  which have a complicated alignment process and inflicts aberrations in DoF. It is important to stress Dove Prism lightly affects the polarization DoF. Consequently, some outputs that ideally should have no intensity are observed to be illuminated. This will impact the quantum state tomography process, leading to a decrease in fidelity. 

\begin{figure}[ht!]
		\centering		\includegraphics[scale=0.45,trim=0cm 0cm 0cm 0cm, clip=true]{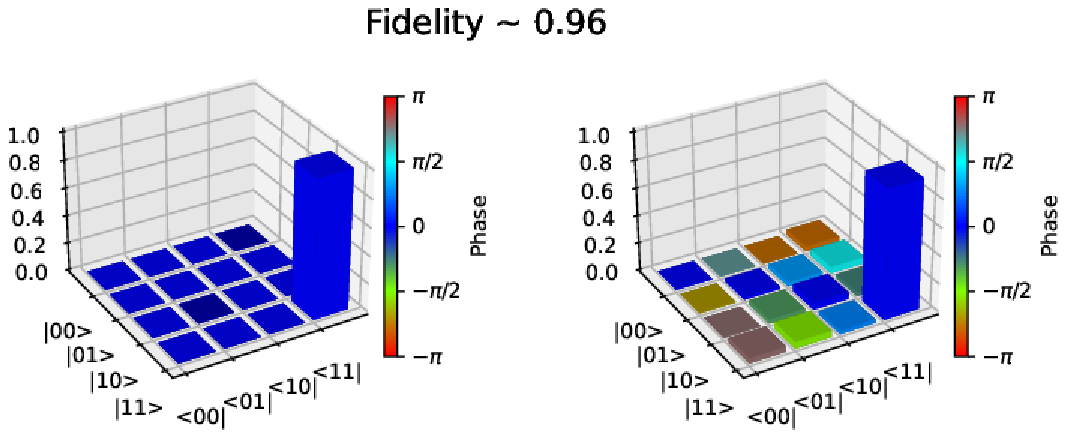}
		\caption{ Theoretical (left) and  Experimental (right) density matrix for the mode $\rho_1$ in the case $c=0$ corresponding to the separable mode $\ket{Vv}$. Fidelity of reconstructed mode is $F=0.96$.}
		\label{d1}
\end{figure}

  Figure \ref{d1} presents the theoretical (left) and experimental (right) density matrix of the mode $c=0$. As we can see, we have a good agreement with theoretical expectations. The Fidelity of the reconstructed mode is $F=0.96$. Mode $\ket{Vv}$ is not hard to prepare, and its fidelity would be closer to unity. This points out the fact that the main question for fidelity in this case is not the mode preparation but the tomography process that requires astigmatic components. All these difficulties discussed above introduce noise in the measurements and impose a limit on the fidelity of the reconstructed density matrix.

\begin{figure}[H]
		\centering
		\includegraphics[scale=0.48,trim=0cm 0cm 0cm 0cm, clip=true]{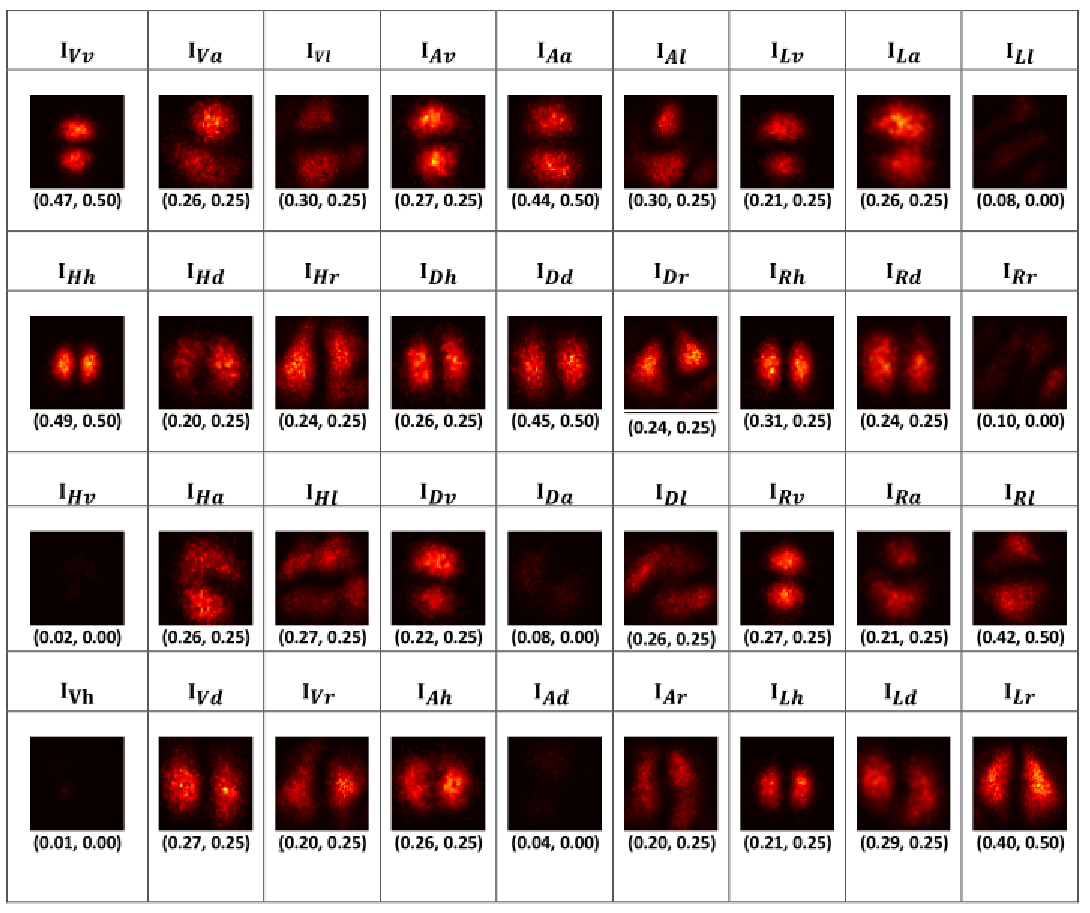}
		\caption{ Recorded images (false color) for maximally nonseparable state $\ket{\phi^+}$ tomographic measurements. Within the parenthesis, we provide the experimental ($I{exp}$) the theoretical ($I{theo}$), and normalized intensities. The pair $(I_{exp}, I_{Theo})$ corresponds to experimental and theoretical normalized values, respectively, for the correspondent basis element.}
		\label{table2}
	\end{figure}

 Let us present the results for the mode with $c=1.00$, that corresponds to the maximally nonseparable pure mode $\ket{\phi^+}$. Figure \ref{table2} presents the recorded image (false color) obtained for the tomographic measurement. The pair $(I_{exp}, I_{Theo})$ corresponds to experimental and theoretical normalized values, respectively, for the correspondent basis element.

\begin{figure}[H]
		\centering
		\includegraphics[scale=0.45,trim=0cm 0cm 0cm 0cm, clip=true]{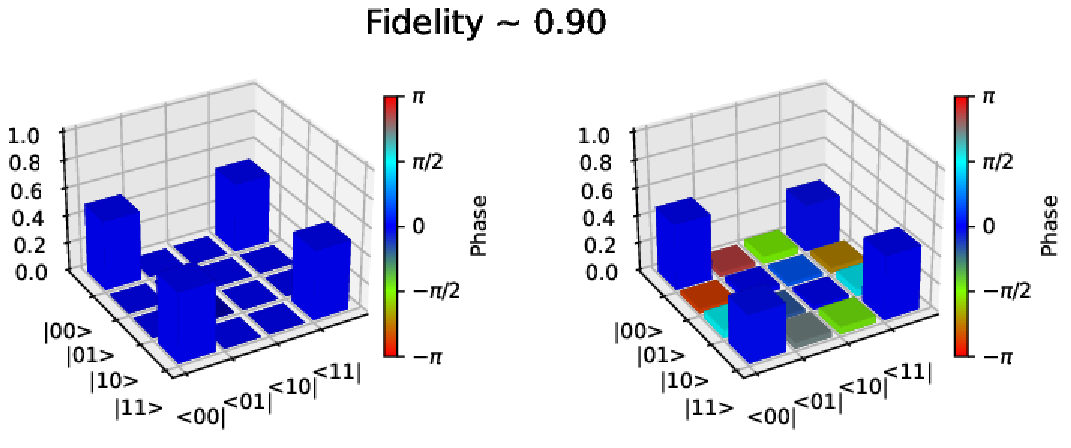}
		\caption{ Theoretical (left) and Experimental (right) density matrix for $\rho_1$ in the case $c=1$, corresponding to the maximally nonseparable mode $\ket{\phi^+}$. The Fidelity for reconstructed mode is $F=0.90$.}
		\label{d2}
	\end{figure}

Theoretical (left) and Experimental (right) density matrix for the mode $\rho_1$ with $c=1$ ($\ket{\phi^+}$) is presented in Figure \ref{d2}. As shown, we have a very good agreement between theory and experiment with Fidelity $F=0.97$. The mode $\ket{\phi^+}$ was prepared with a S-wave plate, that presents high conversion efficiency. We believe the limitation of the Fidelity is due to tomographic measurements.   
  
\begin{figure}[H]
		\centering		\includegraphics[scale=0.48,trim=0cm 0cm 0cm 0cm, clip=true]{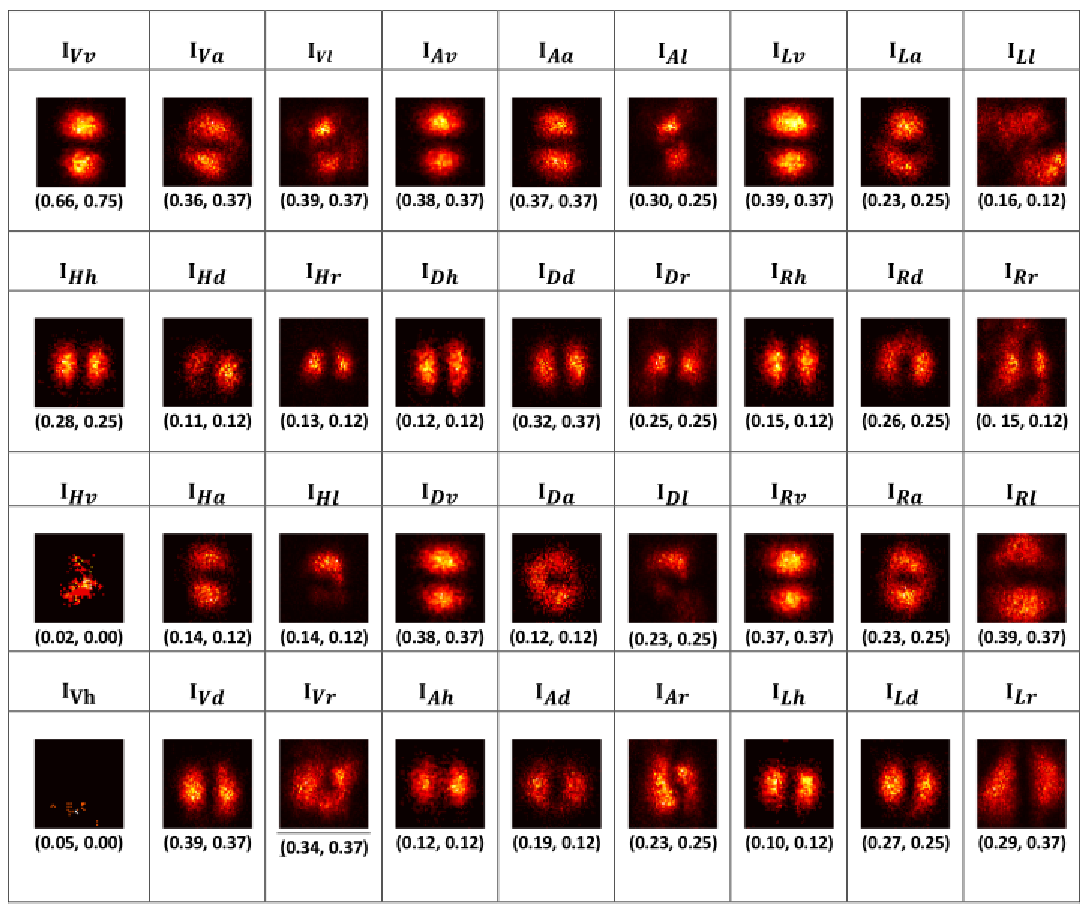}
		\caption{  Recorded images (false color) for $\rho_1$ in the case $c=0.5$, corresponding to a balanced mixed mode for tomographic measurements. Within the parenthesis, we provide the experimental ($I{exp}$) and the theoretical ($I{theo}$) normalized intensities. The pair $(I_{exp}, I_{Theo})$ corresponds to experimental and theoretical normalized values, respectively, for the correspondent basis element.}
		\label{table3}
	\end{figure}

 Let us present the results for $\rho_1$ in the case $c=0.5$, that corresponds to a balanced mixed mode. Figure \ref{table3} present the captured images. Note that clearly we can identify the composition of the intensities of the images of Figures \ref{table1} and \ref{table2} once the intensities coming from each laser is added in the CCD. We can verify in the bottom of each image an excellent agreement between experimental results and theoretical expectation for normalized intensity.

\begin{figure}[H]
		\centering
		\includegraphics[scale=0.45,trim=0cm 0cm 0cm 0cm, clip=true]{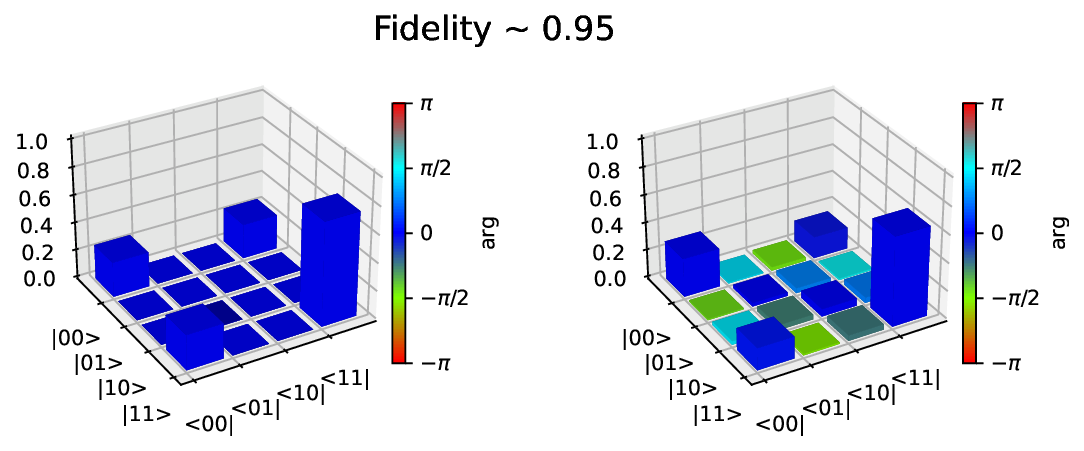}
		\caption{ Theoretical (left) and Experimental (right) density matrix for $\rho_1$ in the case $c=0.5$, corresponding to a balance mixed mode given by Eq.(\ref{eq6}). The Fidelity for reconstructed mode is $F=0.95$.}
		\label{rho1c05}
	\end{figure}

The obtained density matrix for $\rho_1$, given by Eq.(\ref{eq6}), for $c=0.5$  is showed in the Figure \ref{rho1c05}. Theoretical (left) and  Experimental (right) results are in excellent agreement with Fidelity of $0.95$. As observed for previous case, we have a decrease in the coherence terms and a variation in the phase, due the reasons already explained above.

\begin{figure}[H]
		\centering
		\includegraphics[scale=0.40,trim=0cm 0cm 0cm 0cm, clip=true]{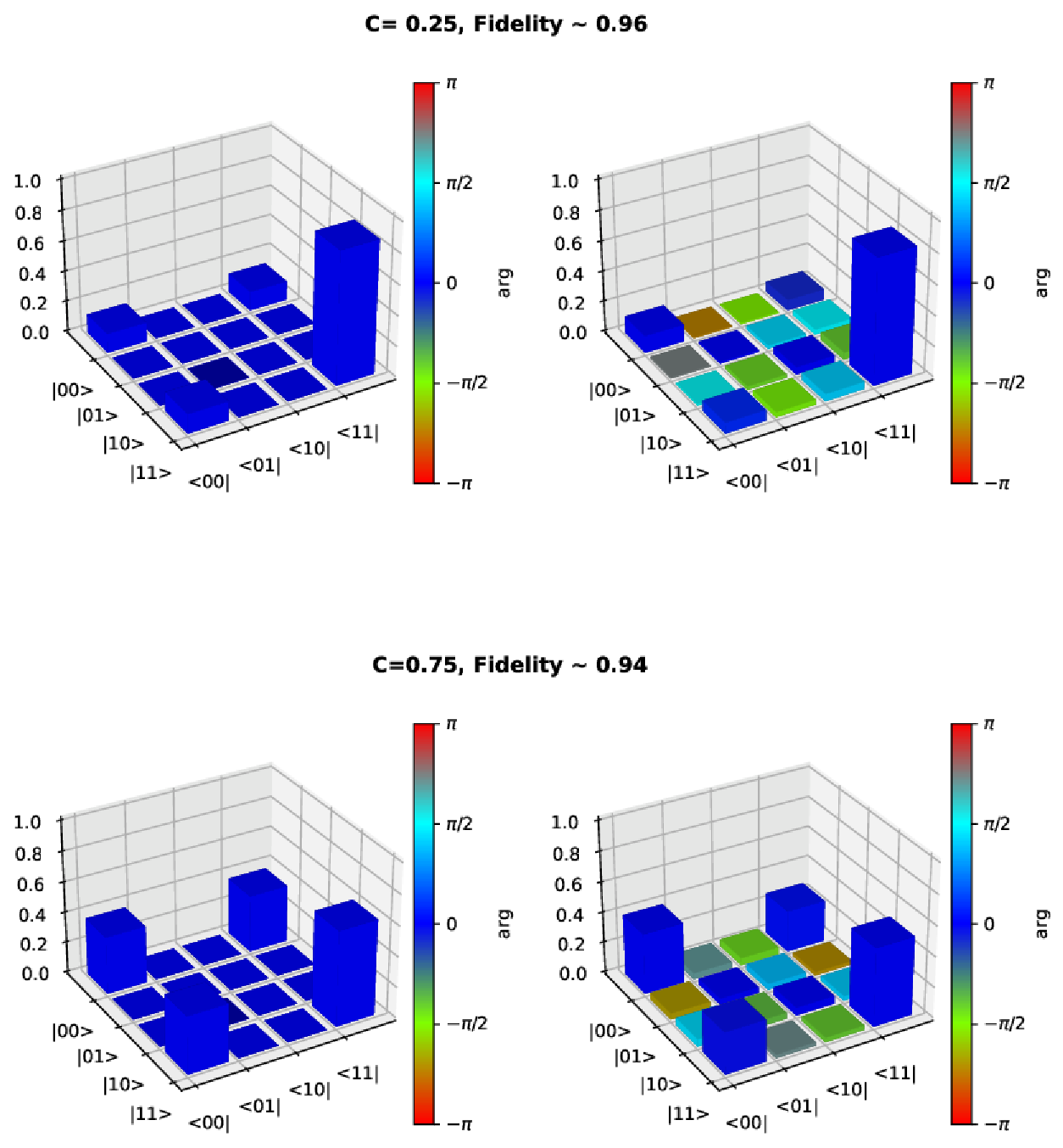}
		\caption{ Density matrices for state $\rho_1$. The weights are represented in top left side. The theoretical matrices are in the left side while the experimental one is on the right.}
		\label{rho1}
	\end{figure}

 Now, let us present the obtained density matrix for the others two investigated mixed modes. Figure \ref{rho1} shows the Theoretical (left) and  Experimental (right) density matrix for the mode $\rho_1$, given by Eq.(\ref{eq6}), for $c=0.25$ and $c=0.75$. The results are in good agreement and the Fidelity is upper than $0.95$.  

\begin{figure}[H]
		\centering
		\includegraphics[scale=0.4,trim=0cm 0cm 0cm 0cm, clip=true]{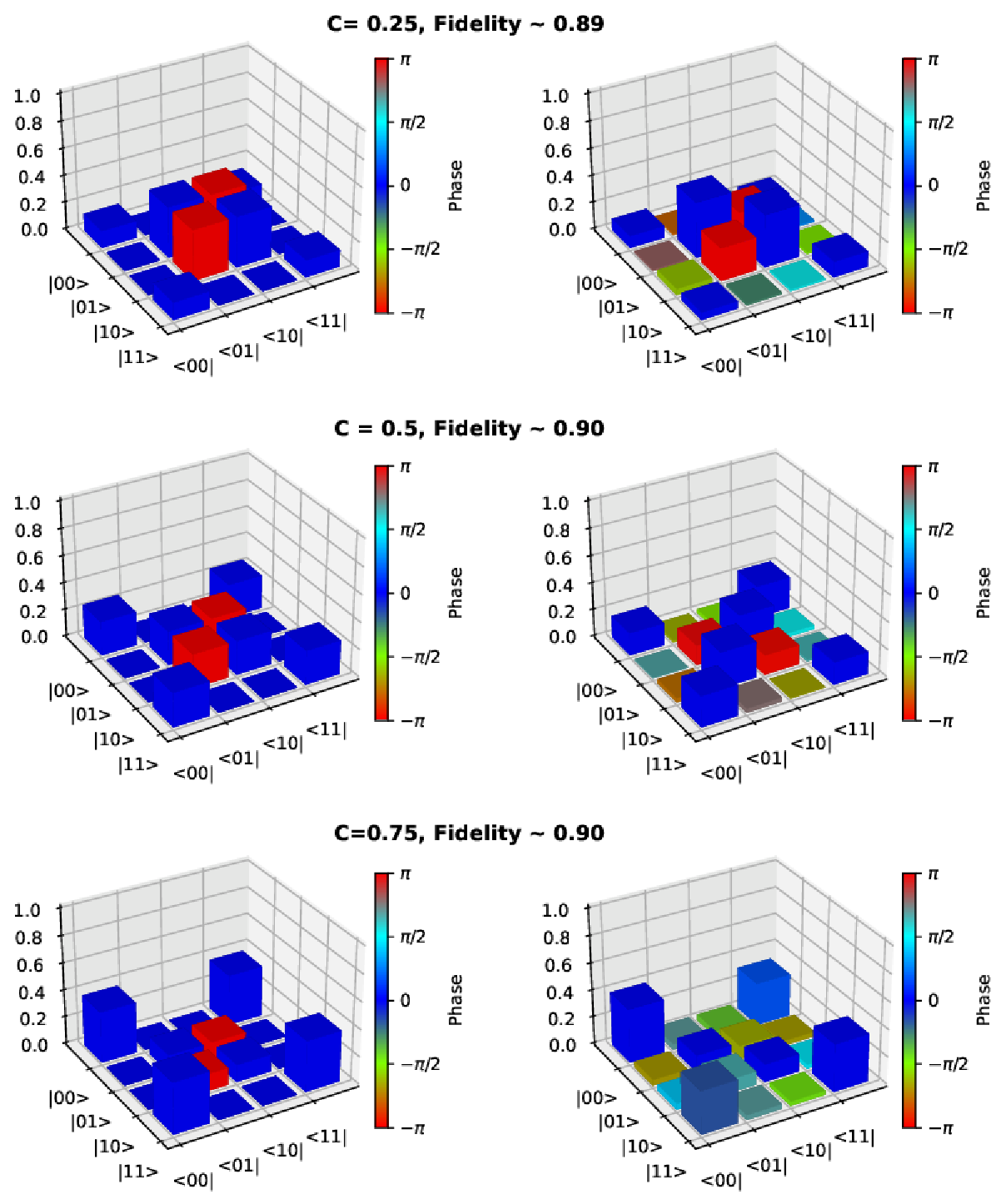}
		\caption{ Density matrices for state $\rho_2$. The weights are represented in top left side. The theoretical matrices are in the left side while the experimental one is on the right.}
		\label{rho2}
	\end{figure}

For the modes $\rho_2$ and $\rho_3$ we present the obtained density matrix.   Figure \ref{rho2} shows the results for the density matrix of  the mode $\rho_2$, given by Eq.(\ref{eq7}), for  $c=0.25$, $c=0.5$, and $c=0.75$. A good agreement between the experimental result (right) and theoretical expectation (left) was observed. Fidelity is around $0.90$.

\begin{figure}[H]
		\centering
		\includegraphics[scale=0.4,trim=0cm 0cm 0cm 0cm, clip=true]{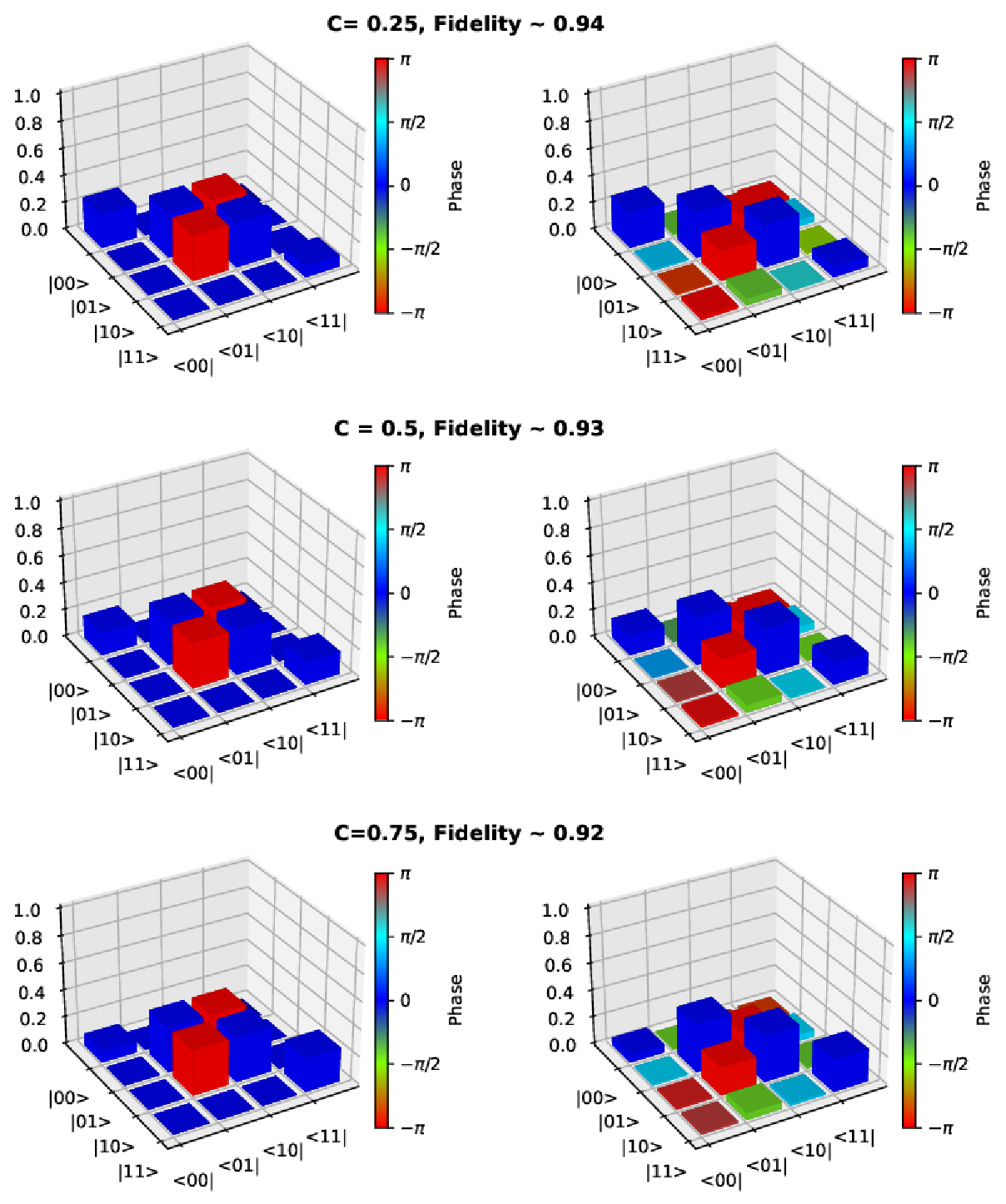}
		\caption{ Density matrices for state $\rho_3$. The weights are represented in the top left side. The theoretical matrices are on the left side while the experimental one is on the right.}
		\label{rho3}
	\end{figure}

Finally, results for the density matrix of the mode $\rho_3$  (Eq.\ref{eq8}) are presented in Figure \ref{rho3}. Again we can see a good agreement between experimental result (right) and theoretical expectation (left) for $c=0.25$, $c=0.5$, and $c=0.75$. Fidelity is higher than $0.90$.

As studied in Section III, limited Fidelity affects the quantum discord calculation. Then, let us present the experimental results for discord.

\subsection{Experimental quantum discord for spin-orbit modes}

After estimating the density matrix by tomography process one can directly compute the QD as presented in Section II-A.

\begin{figure}[H]
		\centering
		\includegraphics[scale=0.4,trim=0cm 0cm 0cm 0cm, clip=true]{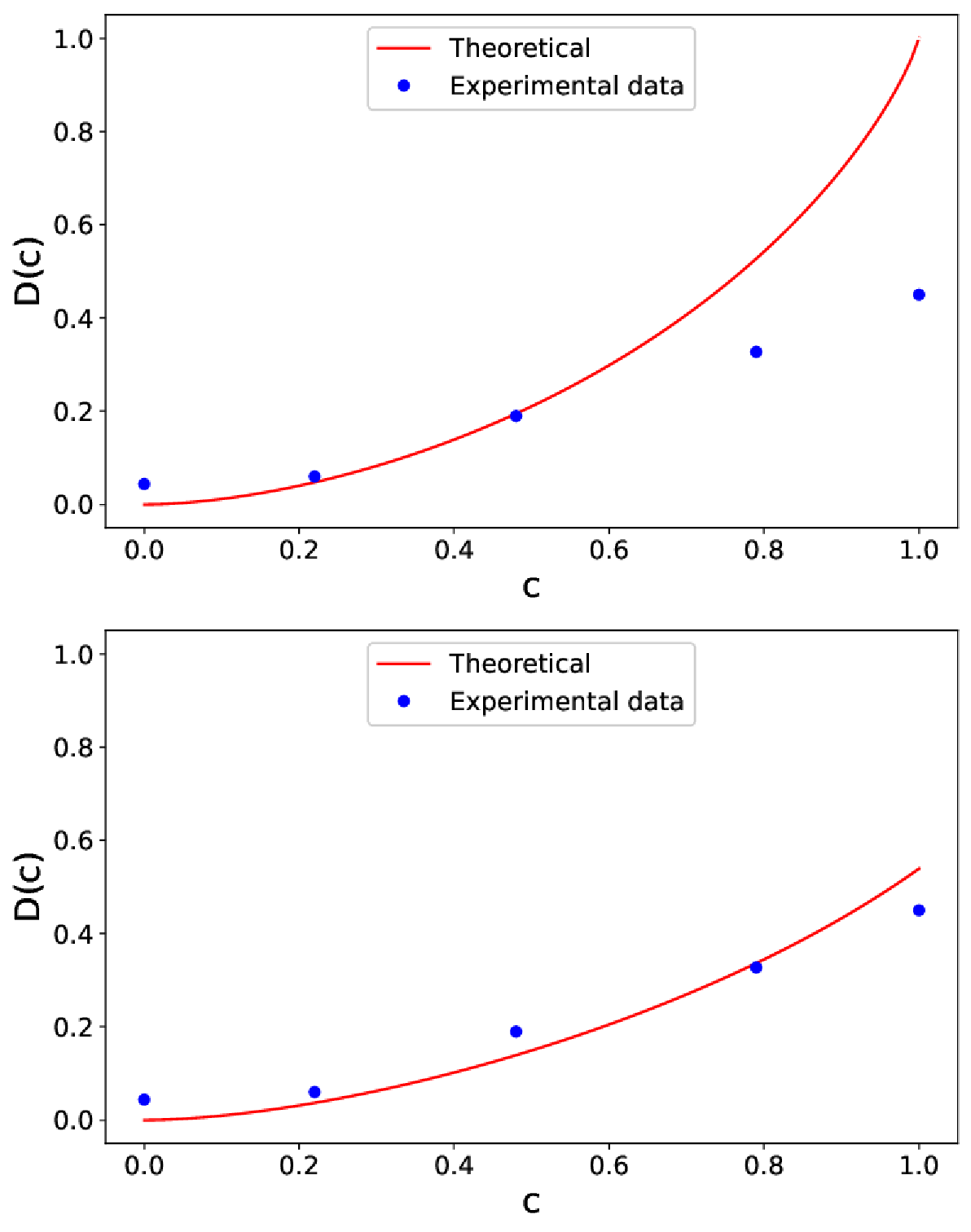}
		\caption{Discord fitting curve for the state $\rho_1$. The ideal curve is represented on top of the figure while the fitted curve is on the bottom. The fitting was done considering the mean fidelity of 0.94.}
		\label{discp1-fit}
	\end{figure}

Figure \ref{discp1-fit} (top) presents the Discord for the mode $\rho_1$ as a function of $c$. The experimental results for $c=0.0;~0.25;~0.5;~0.75;1.0$ are represented by the dots while the solid line is the theoretical expectation. The results agree until $c=0.5$, disclosing exactly as predicted in Section III.  In  Figure \ref{discp1-fit} (down) we show the fitting of experimental dots by the expected theoretical curve with Fidelity $F=0.95$, the mean Fidelity obtained from the fidelity of the five density matrix of the prepared modes of the class $\rho_1$. As can be seen, we have an excellent agreement between our results and the expected discord for the experimental mean Fidelity. 

\begin{figure}[H]
		\centering
		\includegraphics[scale=0.4,trim=0cm 0cm 0cm 0cm, clip=true]{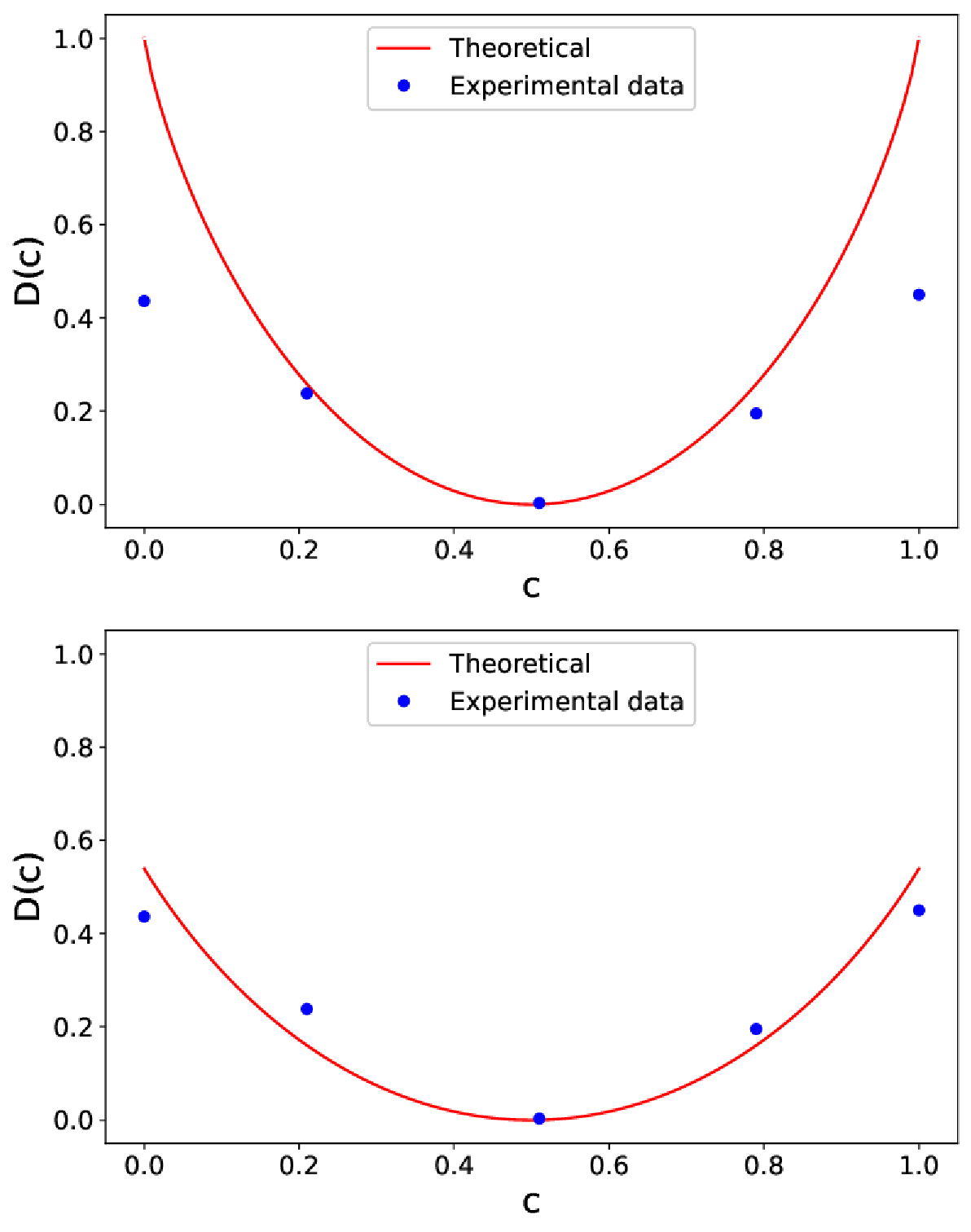}
		\caption{ Discord fitting curve for the state $\rho_2.$he ideal
curve is represented on top of the figure while the fitted curve is
on the bottom. The fitting was done considering the mean
fidelity of 0.95.}
		\label{discp2-fit}
	\end{figure}

The same analysis can be performed for the class of modes $\rho_2$, as shown in Figure \ref{discp2-fit}. In the top image, we can see the experimental results agree with the theoretical expectation for $0.2 < c < 0.8$. In the down image, we can see the fitting of experimental data with the theoretical curve for the mean Fidelity $F=0.90$. These results are also in good agreement with theoretically expected.

\begin{figure}[H]
		\centering		\includegraphics[scale=0.4,trim=0cm 0cm 0cm 0cm, clip=true]{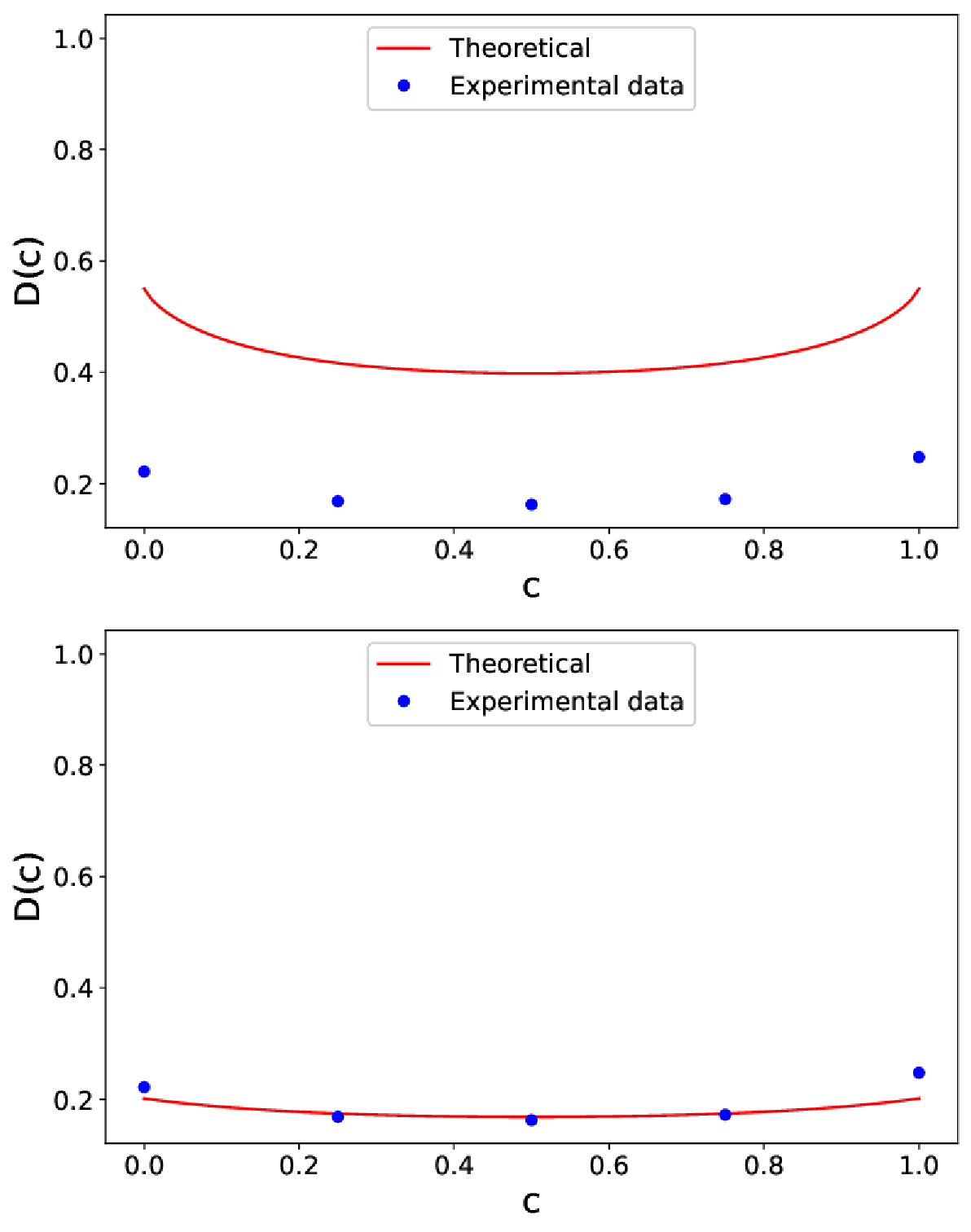}
		\caption{ Discord fitting curve for the state $\rho_3.$ The ideal curve is represented on top of the figure while the fitted curve is
on the bottom. The fitting was done considering the mean
fidelity of 0.95.
}
		\label{discp3-fit}
	\end{figure}

Finally, for $\rho_3$, Figure \ref{discp3-fit} (top) shows similar behavior observed in section III for the class $\rho_3$, while Figure \ref{discp3-fit} (down) shows the fitting of Discord. The experimental mean Fidelity, $F=0.95$, aligns with the results presented in section III. 

This analysis allows us to visualize the effects caused by experimental errors that reduce fidelity and impact directly in the discord calculation. In addition, it's important to emphasize that the large number of steps during discord computation also has a direct impact on the final result. In this way, any small error on the density matrix is propagated on all those steps resulting in the behavior observed. Consequently, even for a high level of fidelity, we observed a substantial deviation in QD value as can be seen in the curves above.

\section{Conclusions}

In summary, we presented the experimental preparation and characterization of the mixed spin-orbit mode proposed in ref.\cite{Xstate}. We experimentally performed two-qubit tomography in polarization and transverse mode degrees of freedom of light and reconstructed the density matrix for three classes of spin-orbit mixed modes. The partial agreement between experimental discord and theoretical expectation can be understood by the strong dependence of the quantum discord on the fidelity of the mixed states. In our case, the lower fidelity is due to several sources of experimental errors impacting directly the measurement outcomes. These errors are primarily attributed to the presence of astigmatic elements that, unfortunately, distort the wavefront, resulting in experimental errors. Despite these issues, we observed a high level of agreement, achieving a fidelity of about 0.90. After getting the density matrices experimentally, we proceeded to calculate the quantum discord. The QD behavior observed, taking into account the experimental errors, agrees with predictions admitting lower fidelity, which is associated with imperfections in experimental realization. As we can see,  the spin-orbit modes constitute an important platform for quantum information processing. Consequently, the limitations related to experimental density matrix reconstruction turn explicit a demand for enhanced methods for measurement and unitary transformation in spin-orbit modes. Indeed, characterizing and manipulate such modes is critical for fully exploring the potential of this platform in quantum information processing.

\begin{acknowledgments}
	We would like to thank financial support from the Brazilian funding agencies Conselho Nacional de Desenvolvimento Cient\'{\i}fico e Tecnol\'ogico (CNPq), 
Funda\c{c}\~ao Carlos Chagas Filho de Amparo \`a Pesquisa do Estado do Rio de Janeiro (FAPERJ), 
Coordena\c{c}\~ao de Aperfei\c{c}oamento de Pessoal de N\'{\i}vel Superior (CAPES) (Finance Code 001), 
and the Brazilian National Institute for Science and Technology of Quantum Information (INCT-IQ).
\end{acknowledgments}
	
	%\nocite{*}
	
{99}

\end{document}